\documentclass[12pt]{article}
\begin{document}
\begin{center}
{\bfseries
Photoabsorption sum rules and quark structure parameters of
hadrons
}
\vskip 5mm

S.B. Gerasimov$^{ \dag}$
\vskip 5mm
{\small
{\it
Bogoliubov Laboratory of Theoretical Physics\\
Joint Institute for Nuclear Research,
141980 Dubna (Moscow region), Russia}
\\
$\dag$ {\it
E-mail: gerasb@thsun1.jinr.ru
}}
\end{center}
\vskip 5mm
\begin{center}
\begin{minipage}{150mm}
\centerline{\bf Abstract}
Following the idea of the quark-hadron duality we present, within the
constituent quark model approach, the relations between different
bremsstrahlung-weihgted integrals of the nucleon resonance
photoexcitation cross sections and correlation functions of the quark dipole
moments in the nucleon ground state. These functions are of interest for
checking detailed quark-configuration structure of the nucleon state
vector. Some applications of this approach in meson sector are made and
the role of meson degrees of freedom in the electromagnetic
baryon observables is briefly discussed.

\end{minipage}
\end{center}
\vskip 10mm

\section{Introduction}

In this paper, we make use the specific form of
notion of "quark-hadron duality" suggested by the correspondence principle
with sum rule technique in nonrelativistic theory of interaction of
radiation with matter and continue
earlier started discussion\cite{Ge69,Ge75} about utility
of a number of sum rules that cannot be derived via
the dispersion-theoretic approach and require more specific
assumptions on validity of asymptotic behaviour of the scattering
amplitudes under consideration.

Sum rules equating integrals over photoabsorption cross-sections to static
parameters of the ground state have long be in the use in nonrelativistic
atomic and nuclear physics (e.g.\cite{Lev}). In relativistic domain, one
should meet problem of the convergence of corresponding integrals and
the necessity to treat all radiative transitions relativistically, i.e.
to include into consideration all higher multipoles in contrast with
atomic and nuclear photoeffect where one can confine oneself by inclusion
the electric dipole amplitudes in the reasonably large energy interval.

Here, our main idea is to merge the duality concept of the Regge-resonance
type and the concept of the parton-hadron duality within the definite
class of sum rules for the photoabsorption cross sections.
These sum rules, at the same time, will qualitatively demonstrate
justifiable common base with the known and much more simple sum rules in
the nonrelativistic domain.

\section{From Regge-duality to quark-hadron duality: The case of
bremsstrahlung-weighted sum rules}

In this section, we propose
and discuss sum rules giving a connection between the valence
quark part of the Dirac (charge) radii of nucleons (mesons)
$<r^2>$ and the photoexcitation
cross sections of the nucleon (or, correspondingly, meson)
resonances. We assume that both the ground and
excited states of hadrons are the bound states of three (or two)
constituent
quarks and, tentatively, that the electroweak coupling constants of
"constituent" and "current" quarks are the same (i.e. obtained via the
minimal coupling principle).

It follows then, that the usual current algebra
relations are valid for the currents constructed of the constituent quark
field operators and, at the same time, one can to replace the sum over
complete set of the hadron states in the dispersion integrals by the sum
over all resonance states constructed, for example,
of three constituent quarks and having the needed quantum numbers.

The first problem to deal with is the convergence of integrals, containing
the moments of considered cross-sections. We rely here on
the known idea of semi-local duality between the description of
a given amplitude behaviour at high energies via the leading
non-vacuum Regge
trajectory exchanges in the t-channel and the sum of the s-channel
resonances\cite{DHS68}.
The vacuum Pomeron exchanges should then be associated with the
non-resonance s-channel amplitudes\cite{HaFr68}.

Accepting that kind of duality between the
resonance photoexcitation cross-sections and contributions of the positive
charge-parity Regge exchanges, the $f$- and $a_2$-exchange trajectories,
we conclude that the bremsstrahlug-weighted integrals of the total
photoproduction cross-section should be convergent ones.

Now, we consider first the nucleon case.
Our basic idea consists in relating the
electric dipole moment correlator and the mean-squared radii operators
sandwiched by the nucleon state vectors in the "infinite - momentum frame"

\begin{eqnarray}
&& 2<\hat{D^2}>_{N} - <\hat{D^2}>_{P} +
8<\hat{D_{S}^2}>_{P(N)} = 2<\hat{r_{1}^2}>_{N} + <\hat{r_{1}^2}>_{P}
\end{eqnarray}
where
\begin{eqnarray}
&& \hat{D} = \int \vec{x} \hat{\rho}(\vec{x})d^3x = \sum_{j=1}^{3}Q_q(j)\vec{\xi_{j}}\\
&& \hat{r_{1}^2} = \int \vec{x}^2 \hat{\rho}(\vec{x})d^3x = \sum_{j=1}^{3}Q_q(j)\vec{\xi_{j}}^2
\end{eqnarray}
$Q_q(j)$ and $\vec{\xi_{j}}$ - are the electric charges and configuration
variables of presumably point-like constituent quarks
in the infinite-momentum frame while the electric charge density operator
$\hat\rho={\hat\rho}^{S}+{\hat\rho}^{V}$ is a sum of the isoscalar and
isovector parts.The relation (13) comes from the following parametrization
of the matrix elements \\
\begin{eqnarray}
&& <r_{1}^2>_{P} = \frac{4}{3}\alpha - \frac{1}{3}\beta \\
&& <r_{1}^2>_{N} = -\frac{2}{3}\alpha + \frac{2}{3}\beta \\
&& <\hat{D^2}>_{P} = \frac{8}{9}\alpha + \frac{1}{9}\beta + \frac{8}{9}\gamma - \frac{8}{9}\delta\\
&& <\hat{D^2}>_{N} = \frac{2}{9}\alpha + \frac{4}{9}\beta + \frac{2}{9}\gamma - \frac{8}{9}\delta\\
&& <\hat{D_{S}^2}>_{P,N} = \frac{1}{36}( 2\alpha + \beta + 2\gamma + 4\delta )\\
\end{eqnarray}

where $<\vec\xi_{1}^{2}>=<\vec\xi_{2}^{2}>=\alpha$,$<\vec\xi_{3}^{2}>=\beta$,
$<\vec\xi_{1}\cdot\vec\xi_{2}>=\gamma$,$<\vec\xi_{1}\cdot\vec\xi_{3}>=<\vec\xi_{2}\cdot\vec\xi_{2}>$\\
$\mbox{}=\delta$
the indices "1" and "2" refer to the like quarks (i.e. to the $u(d)$-
quarks inside proton (neutron)), and the "3" - to the odd quark.
The matrix elements over the proton and neutron wave function are only
assumed to obey to the charge symmetry
relations.

The subsequent procedure is a standard technique of sum rule derivation
within the framework of the dipole moment algebra at
the "$p_z \to \infty$" - frame\cite{CR66,Go66}.

However, we attribute a new meaning to all appearing quantities, namely, all
the cross sections are understood as the nucleon resonance excitation
cross sections,all radii $<r_1^2>_{p,n}$ as the constituent quark
distribution radii $<r_1^2>_{p,n}^{b}$, not including the sea quark and/or
meson current effects.Further, we will replace the intermediate
one-nucleon contributions proportional to the anomalous magnetic moments
of nucleons, by the corresponding integrals entering in the anomalous
magnetic sum rules.One can then to get

\begin{eqnarray}
&& \frac{4}{3}\pi^2 \alpha <\hat{D^2}>_{P(N)} = J_{p}^{\gamma P(N)} = J_{p}^{\gamma P(N)}(\frac{1}{2}) + J_{p}^{\gamma P(N)}(\frac{3}{2})\\
&& \frac{4}{3}\pi^2 \alpha <\hat{r_{1}^2}>_{N}^{b} = J_{a}^{\gamma P}(\frac{3}{2}) - J_{p}^{\gamma P}(\frac{1}{2}) + 4J_{p}^{S}(\frac{1}{2})
\end{eqnarray}
where
\begin{eqnarray}
&& J_{p,a}^{V(S)}(I) = \int_{\omega_{thr}}^{\infty}\frac{d\omega}{\omega}\sigma_{p,a}(\gamma^{V(S)}N \to N^{\star}(I))
\end{eqnarray}

the $\sigma_{p,a}$ refers to the interaction cross section of the
polarized "isovector"("isoscalar")  $(\gamma^{V,S})$ photons and polarized
nucleons with parallel (or antiparallel) spins.

The relativistic dipole-moment fluctuation sum rule,Eq.(10),
has been checked\cite{Ge75}
in a number of the field-theoretical models, while
the limiting case of
Eq.(11) with the assumption of fully symmetrical quark distributions in
nucleons,i.e. when
$\alpha=\beta$ and $\gamma=\delta$, giving $<r_1^2>_{n}^{b}=0$, has been
discussed in\cite{Ge69} .

The calculations have shown that
the Dirac charge radius of the neutron is indeed a small quantity,especially
if the model amplitudes are used taking into account the relativistic
corrections and the effects of the mixing of the basis
$SU(6) \otimes O(3)$- configurations in the ground and excited
wave functions\cite{Clo}.
However,if the use is made of the phenomenological amplitudes (especially
enhanced, as compared with the quark model calculations, of
the $\Delta(1232)$ excitation amplitude), the right-hand side of Eq.(11)
acquire larger positive value

\begin{eqnarray}
&& \frac{4}{3}\pi^2 \alpha <\hat{r_{1}^2}>_{N}^{b} \simeq 0;~( 34 \mu b )
\end{eqnarray}

where the first value corresponds to the model\cite{Clo}
and the second one - to
phenomenological $\gamma N \to N^{\star}(I=1/2,3/2)$-transition amplitudes,
respectively\cite{PDG}.

The most precise determination of the neutron charge radius follows from
the neutron-electron scattering lengths measured in the thermal neutron
scattering off the inert gases ( Ne,Ar,Kr,Xe), or metal ( W,Pb,Bi ) atoms.
The dominant contribution to the (Sachs) charge radius of the neutron\\
\begin{eqnarray}
&& <\hat{r_{ch}^2}>_{N} = <\hat{r_{1}^2}>_{N} + \frac{3\kappa_{N}}{2m_{N}^2}
\end{eqnarray}
is due to the second ( Foldy) term, proportional to $\kappa_{N}=-1.913$ n.m.
The experimental uncertainties ascribed to a measured values of $<r_{ch}^2>_{N}$
enable to extract a very small contribution of
the Dirac (charge) radius $<r_{1}^2>_{N}$ from Eq.(14)\cite{Ko95}.

The sum rules for mesons have been written down as well
\begin{equation}
\frac{4}{3} \pi^2 \alpha <r^2>_{\pi^{+}} = J_{tot}(\gamma \pi^{+})
+ \frac{4}{5} J_{tot}(\gamma \pi^{o}),
\end{equation}
\begin{equation}
\frac{4}{3} \pi^2 \alpha <r^2>_{K^{+}} = J_{tot}(\gamma K^{+})
+ 2J_{tot}(\gamma K^{o}).
\end{equation}
and have been checked analytically in the lowest
perturbation order in the meson-quark coupling constant\cite{Ge75}.
Their phenomenological consequences have been discussed in more detail in
\cite{Ge79}.

In what follows we rather turn to sum rules for meson
resonances in photon-photon collisions.
Their evident advantage is in that the partonic "wave function" of the
photon follows from
the lowest order
perturbation calculation in the fine coupling
constant and standard local coupling in the photon-charged-parton
vertices.  Furthermore, it can be shown that
choosing different polarizations of
colliding photons, one can obtain the linear combination of certain
$\gamma \gamma \to q\bar q $ cross-sections
that will dominantly collect
the $q\bar q$- states with definite spin-parity, hence, by the adopted
quark-hadron duality, the meson resonances with the same quantum numbers.
In particular, the combinations of the integrals over the
bremsstrahlung-weighted and polarized
$\gamma \gamma \to q\bar q $ cross-sections,
$I_{\bot} - (1/2)I_{p}, I_{\|} - (1/2)I_{p}, I_{p}$ will be related with
low mass meson resonances having spatial quantum numbers ($J^{PC} =
0^{-+},~~ 0^{++},~~2^{++}$), correspondingly. The
$\gamma \gamma $ - cross-sections
$\sigma^{\gamma \gamma} _{\bot (||)}$
(and the integrals thereof) refer to
plane-polarized photons with the perpendicular (parallel)
polarizations, and the $\sigma^{\gamma \gamma} _{p} $  corresponds to
circularly polarized photons with parallel spins.

Evaluating cross-sections and elementary integrals for the
experimentally best known case of pseudoscalar mesons we get the
sum rule for radiative widths of the $\pi^{o}, \eta $\\
 and $\eta^{\prime}$~~mesons

\begin{eqnarray}
\frac{1}{16\pi^2}\int_{thr}^\infty
\frac{ds}{s}(\sigma_{\bot}^{\gamma \gamma}(s)-\frac{1}{2}\sigma_{p}^
{\gamma \gamma}(s)) = \frac{\alpha^2}{24\pi^2 m_{q}^2} & = &
\sum_{i=\pi,\eta,\eta^{\prime}\ldots}(\frac{\Gamma(i \to
\gamma \gamma)}{m_{i}^3})\\
(12.3 \cdot 10^{-6} \mbox{GeV $^{-2}$}) & \simeq & (10.9\cdot10^{-6}
\mbox{GeV $^{-2}$}) \nonumber
\end{eqnarray}
The value $m_q \simeq 240$~MeV
%$g_{\pi qq}$
used in this calculation
have been taken from\cite{Ge79}~
(in Eq.(17), we have taken approximately $m_{u,d} \simeq m_s$ which is
within accuracy of our calculation).  Close numerical values of the
integral and sum of known radiative widths\cite{PDG} leave little place for
contribution of higher mass pseudoscalar radial excitations and give good
evidence for the relevance of quark- hadron duality in the considered
context.

\section{The pionic "dressing" of hadrons: examples and problems}

In this section we discuss briefly the role of nonvalence degrees of
freedom ( the nucleon sea partons and/or peripheral meson currents )
in parametrization and description of hadron magnetic moments, both
diagonal and non-diagonal, including the $N\Delta
\gamma$-transition moment.
Earlier we have considered\cite{Ge95} a number of
consequences of sum rules for the static,
electroweak characteristics of baryons following from the theory of broken
internal symmetries and common features of the quark models
including relativistic effects and corrections due to
nonvalence degrees of freedom -- the sea partons and/or the meson clouds
at the periphery of baryons.\\
Now, we list some consequences of the obtained sum rules.
The numerical relevance of adopted parametrization is seen from results
enabling even  to estimate from one of obtained sum rules,
namely,
\begin{eqnarray}
&&(\Sigma^+ - \Sigma^-) (\Sigma^+ + \Sigma^- - 6\Lambda + 2\Xi^0 + 2\Xi^-)\nonumber\\
&&- (\Xi^0 - \Xi^-) (\Sigma^+ + \Sigma^- + 6\Lambda - 4\Xi^0 - 4\Xi^-) = 0.
\end{eqnarray}
the necessary effect of the isospin-violating $\Sigma^{o} \Lambda$-mixing.
By definition, the $\Lambda$--value
entering into Eq.(18)
should be "refined" from the
electromagnetic $\Lambda\Sigma^0$--mixing affecting  $\mu(\Lambda)_{exp}$.
Hence, the numerical value of $\Lambda$, extracted from Eq.(18),
can be used to determine the $\Lambda\Sigma^0$--mixing angle through the
relation
\begin{eqnarray}
\sin \theta_{\Lambda\Sigma} \simeq \theta_{\Lambda\Sigma} =
{\Lambda -\Lambda_{exp} \over 2\mu(\Lambda\Sigma)} = (1.43 \pm 0.31) 10^{-2}
\end{eqnarray}
in accord with the independent estimate of $\theta_{\Lambda\Sigma}$ from
the electromagnetic mass-splitting sum rule \cite{DvH}.
Owing to interaction of the $u$-- and $d$-- quarks with charged
pions the "magnetic anomaly" is developing, i.e. $u/d=-1.80\pm 0.02 \not=
Q_u/Q_d = -2.$ Evaluation of the lowest order quark--pion loop diagrams
gives \cite{Ge95}: $u/d=(Q_u +\kappa_u)/(Q_d+\kappa_d)=-1.77$, where
$\kappa_q$ is the quark anomalous magnetic moment in natural units.

Of course, this approach is free of a problem raised by Lipkin\cite{Lip99}
and concerning the ratio $R_{\Sigma / \Lambda}$
of magnetic moments of $\Sigma$- and $\Lambda$- hyperons.
With the parameters
$u/d=-1.80$ and $\alpha_{D}=(D/(F+D))_{mag}=0.58$, defined without
including in fit the $\Lambda$-hyperon magnetic moment, we obtain
\begin{equation}
R_{\Sigma / \Lambda} = \frac{\Sigma^{+}+2\Sigma^{-}}{\Lambda}
= -.27~~ (\mbox{\it vs}~~ -.23~ \cite{PDG})
\end{equation}
while in the standard nonrelativistic quark model
without inclusion of non-valence d.o.f. this ratio would equals $-1$.

The meson--baryon universality of the quark characteristics,
suggested long ago \cite{Ge65}, is confirmed by
the calculation of the ratio of $K^*$ radiative widths
\begin{eqnarray}
{\Gamma(K^{*+}\to K^+\gamma)\over \Gamma(K^{*+}\to K^+\gamma)} =
\left({u/d} + {s/d}\over 1+ {s/d} \right)^2 =0.42\pm 0.03 \
(\mbox{\it vs} \ 0.44\pm 0.06 \mbox{\cite{PDG}}).
\label{e16}\end{eqnarray}

The experimentally interesting quantities $\mu(\Delta^+ P) = \mu(\Delta^0 N)$
and $\mu(\Sigma^{*0}\Lambda)$ are affected by the exchange current
contributions and for their estimation we need additional assumptions.
We use the analogy with the one--pion--exchange current, well--known in
nuclear physics, to assume for the exchange magnetic moment operator
\begin{eqnarray}
\hat\mu_{exch} =\sum\limits_{i<j} [\vec\sigma_i \times \vec\sigma_j]_3
[\vec\tau_i \times \vec\tau_j]_3 f(r_{ij}),
\label{e17}\end{eqnarray}
where $f(r_{ij})$ is a unspecified function of the interquark distances,
$\vec\sigma_i (\vec\tau_i)$ are spin (isospin) operators of quarks.
Calculating the matrix elements of $\hat\mu_{exch}$ between the baryon wave
functions, belonging to the 56--plet of $SU(6)$, one can find
\begin{eqnarray}
&& C(P)={1\over \sqrt{2}} C(\Delta^+P) = \sqrt{3} C(\Lambda\Sigma),\label{e18}\\
&& \mu(\Delta^+P)_{\underline 56} = {1\over \sqrt{2}} \left(P - N +{1\over
3} (P+N) {1-u/d\over 1+ u/d}\right).
\end{eqnarray}
where
Eq.(24) may serve as a generalization of the well--known $SU(6)$--relation
\cite{BPL64}.

We list below the limiting relations following
from the neglect of the meson degrees of freedom
\begin{eqnarray}
&& \Sigma^+ [\Sigma^-] = P[-P-N] +
(\Lambda - {N\over 2})( 1+{{2N}\over P}),\\
%\label{e20}
&& \Xi^0[\Xi^-] = N[-P-N] + 2(\Lambda -
{N\over 2})( 1+ {N\over {2P}}),\\
%\label{e21}
&& \mu(\Lambda\Sigma) = -{\sqrt{3}\over 2}N.
\end{eqnarray}
The numerical values of magnetic moments following from this assumption
coincide almost identically with the
results of the $SU(6)$--based NRQM taking account of the $SU(3)$ breaking
due to the quark--mass differences \cite{Ge65}. We stress, however, that no
NR assumption or explicit $SU(6)$-wave function are used this time.
The ratio $\alpha_{D}=D/{(F+D} = .61$ in this case and it is definitely
less than
$ \alpha_{D}=.58$,~ when non-valence degrees of freedom are
included\cite{Ge95}.
This is demonstrating a substantial influence of the nonvalence
degrees of freedom on this important parameter.

\section{Concluding remarks}
The empirical spectrum of mesons and baryons suggests that hadrons
are largely composed of the spin-1/2 constituent quarks confined to
 $q \bar q$ and $qqq$ systems. Naturally, one needs to understand these
relevant degrees of freedom, their effective structure parameters
and forces acting between them to reach the understanding of QCD in the
confining regime.
The constituent quark model is a useful descriptive tool to systematize the
phenomenology of the resonance physics. The photon-hadron inelastic
reactions in the resonance region have an additional attractiveness
in that many phenomenologically successful ingredients of the description
of static electromagnetic properties of hadrons, forming one of
cornerstones of the constituent quark model itself, can be invoked and
included in these processes. Despite many successes the fundamental
question of the connection between the three (for baryons)
( or the $q\bar q$-pair, for mesons) "spectroscopic" quarks,
bearing the quantum numbers of a given hadron, and the infinite
number of the "current" (or fundamental) quarks required,
$\it e.g.$, by the deep-inelastic lepton-hadron scattering, is still
a problem.
One can hope that the soon available precise
experimental data on the resonance photoexcitation amplitudes will
provide desirable constraints on the ground state
quark-configuration structure and the effective coupling constants of
corresponding constituent quarks.
Tentatively, one can conclude that,
for mesons, the abovementioned application of the developed approach
to derivation
of the $\gamma \gamma$- sum rule looks very encouraging. As far as
it is the most
clean test of our approach, its extension for radiative decays
of higher spin meson resonances appears very desirable and interesting.

For the nucleon and nucleon resonances, the situation looks intriguing in
that the $SU(6)$- value
 of the$\Delta N$ - transition magnetic moment
(remaining also in the $N_{c} \to \infty$-approach)
turned out intact to
the included
pionic and relativistic corrections and, probably, may indicate to
the presence of significant "live" meson component in the Fock-state vector
of this resonance. In this respect, the ongoing and forthcoming
detailed investigation of higher nucleon resonance may also bring
interesting surprises of the same sort. The pertinent generalization
of sum rules taking into account these additional degrees of freedom is
still to be done.

\section*{Acknowledgments}
This work was supported in part by
the grant of Plenipotentiary Representative of Slovak Republic in JINR.
The author would like to express his gratitude to organizers of
the conference for warm hospitality extended to him.

\end{document}